\documentclass[letterpaper,11pt]{emulateapj}
\usepackage{epsfig}
\usepackage{natbib}

\begin{document}
\newcommand{\kms}{km~s$^{-1}$}
\newcommand{\Msun}{M_{\odot}}
\newcommand{\Lsun}{L_{\odot}}
\newcommand{\ML}{M_{\odot}/L_{\odot}}
\newcommand{\etal}{{et al.}\ }
\newcommand{\hhh}{h_{100}}
\newcommand{\hsq}{h_{100}^{-2}}
\newcommand{\tn}{\tablenotemark}
\newcommand{\mdot}{\dot{M}}
\newcommand{\p}{^\prime}
\newcommand{\kmsMpc}{km~s$^{-1}$~Mpc$^{-1}$}

\title{The Extragalactic Distance Database}

\author{R. Brent Tully,}
\affil{Institute for Astronomy, University of Hawaii, 2680 Woodlawn Drive, Honolulu, HI 96822}


\author{Luca Rizzi}
\affil{United Kingdom Infrared Telescope, 660 N. A'ohoku Pl, Hilo, HI 96720}


\author{Edward J. Shaya}
\affil{Department of Astronomy, University of Maryland, College Park, MD 20742}


\author{H\'el\`ene M. Courtois}
\affil{Universit\'e Lyon 1, CNRS/IN2P3/INSU, Institut de Physique Nucl\'eaire Lyon, France and Institute for Astronomy, University of Hawaii, Honolulu HI 96822}

\author{Dmitry  I. Makarov}
\affil{Special Astrophysical Observatory, Nizhniy Arkhyz, Karachai-Cherkessia 369167, Russia and Isaac Newton Institute of Chile, SAO Branch, Russia}

\and

\author{Bradley A. Jacobs}
\affil{Institute for Astronomy, University of Hawaii, 2680 Woodlawn Drive, Honolulu, HI 96822}

\begin{abstract}
A database can be accessed on the web at http://edd.ifa.hawaii.edu that was developed to promote access to information related to galaxy distances.  The database has three functional components.  First, tables from many literature sources have been gathered and enhanced with links through a distinct galaxy naming convention.  Second, comparisons of results both at the levels of parameters and of techniques have begun and are continuing, leading to increasing homogeneity and consistency of distance measurements.  Third,  new material are presented arising from ongoing observational programs at the University of Hawaii 2.2m telescope, radio telescopes at Green Bank, Arecibo, and Parkes and with Hubble Space Telescope.  This new observational material is made available in tandem with related material drawn from archives and passed through common analysis pipelines. 
\end{abstract}


\section{Introduction}

The number of galaxies with measured distances is approaching or surpasses $10^4$.  There are of order 10 useful distance-estimator methodologies.  Distances are necessary to constrain the extragalactic distance scale characterized by H$_0$, the Hubble Constant, and to transform observed luminosity, mass, and dimension parameters into intrinsic values.  At a deeper level of complexity, distances are needed to probe departures from the cosmic expansion, to determine the line-of-sight component of `peculiar velocities'.  These motions are thought to arise from variations in the gravity field as a consequence of the distribution of matter, most of it dark.  With the new wealth of data it may be possible to learn a lot more about the expansion of the universe and the distribution of matter.  However, can we trust the quality and homogeneity of the information that is becoming available?

The first order purpose of the Extragalactic Distance Database (EDD) is to assemble data relevant to the determination of the distances of galaxies in one place, then inter compare these data to check for compatibility or not, and finally to generate weighted best estimates of distances to individual galaxies and groups of galaxies all on a common zero point scale.   The second order purpose is to facilitate studies of the large scale distribution and motions of galaxies toward the goals of developing a three-dimensional map of the distribution of matter in the local universe and developing reconstructions of the formation process.  This paper provides an overview of the contents of EDD which can be accessed at http://edd.ifa.hawaii.edu.

\section{The First Page}

The EDD home page provides a brief overview of the contents of the database and provides access to descriptive material that will be updated.  Pressing the `next' button takes one to the `First Page' to be confronted by panels identifying catalogs (presently 58 and counting). Many of these are extracted from the literature, sometimes involving links between two or more original tables.  Some of the catalogs are built from the synthesis of material from a variety of sources.  Other catalogs provide access to graphical material, some obtained by the individuals involved with the database and much drawn from telescope archives and presented here after analysis through our pipelines.

The panels are presented in blocks (currently 7).  Each block contains a group of similar catalogs.
The broad categories are: (1) {\it Redshift Catalogs}: these are compendia of the general properties of galaxies with a variety of selection and completion characteristics; (2) {\it Summary Distances}: a gathering together of distance information from multiple sources; (3) {\it Miscellaneous Distances}: information from important sources but using methods with too few catalogs to command their own block; (4) {\it Photometry}: catalogs of galaxy photometry; (5) {\it HI Linewidths}: sources of neutral Hydrogen profile information; (6) {\it Optical Linewidths}: catalogs of rotation rate information obtained from optical spectroscopy; (7) {\it Fundamental Plane}: compilations of data related to use of this important method.  

EDD is clearly unbalanced in its representation of different distance methodologies.  There is a great deal of material related to the correlation between galaxy luminosities and rotation rates  \citep{1977A&A....54..661T} because we calculate distances with this method from the base material of photometry and spectral information.  With other techniques we accept results as published, except potentially with zero-points.

In the past it has been difficult to compare distance measurements between sources.  A basic problem has been simply the confusion of galaxy names and the uneven quality of galaxy coordinates.  And it does no good to just have the available estimates of distance to a specific galaxy without having a basis for evaluating the merits of each estimate and an understanding of their zero point scaling.

A defining feature of EDD is that every entry in every catalog is linked through a unique naming convention.  Each galaxy in every catalog is identified by the Principal Galaxies Catalog (PGC) number given it in the Third Reference Catalogue \citep{1991trcb.book.....D} or the continuance in LEDA, the Lyon Extragalactic Database\footnote{http://leda.univ-lyon1.fr}  \citep{2003A&A...412...45P}.  In a given catalog, a galaxy can only appear once.  EDD can be used to link information between catalogs.  If, in the extreme, one wanted all the information in EDD regarding a particular galaxy, one could have all the tabular information on one very long row of output.  There are now 1500 elements of (often redundant) information.

As a consequence especially of two of the large redshift catalogs, EDD now contains data on $\sim 50,000$ galaxies.  The discussion now turns to the content of the distinct blocks.

\subsection{Redshift Catalogs}

The first catalog one encounters is entitled {\bf LEDA}   \citep{2003A&A...412...45P}  and it is a special case.  The entire {\it LEDA} database now includes in excess of 4 million galaxies.  Our distillation includes only those objects that enter EDD through one of the other catalogs.  If a new galaxy enters the database through a new catalog then we add the corresponding entry to the {\it LEDA} file.  Consequently, the parameters provided by {\it LEDA} are available for all galaxies in the database, from whatever catalog.  The number of entries in the {\it LEDA} file indicated in the {\it LEDA} selection box is the number of discrete galaxies currently in EDD.  {\it LEDA}, in addition to providing the PGC name used for intra-catalog linkages, is a prime source for coordinates.  It also provides a reasonably uniform identification of morphological types and a useful cross-reference for names.

In all the catalogs, it will be seen that the PGC number in the first column is highlighted blue.  A mouse click brings up a Digital Sky Survey image of the galaxy drawn from the {\it LEDA} site.

The database includes two large redshift catalogs: the {\bf 2MASS K $<$ 11.25} redshift survey,
provided in advance of publication by J. Huchra, and a catalog called {\bf V8k} which includes all galaxies with known redshifts, excluding recent multi-object spectrograph results, within a cube extending 8,000~\kms\ on the cardinal axes from our position.  These catalogs are used to construct maps of the distribution of galaxies and are the bases of reconstructions.   The 2MASS Redshift Survey provides the most complete rigorously defined  all-sky sample of the local universe.  2MASS: the Two Micron All Sky Survey Extended Source Catalog has been described by  \citet{2000AJ....119.2498J} and the properties of the redshift survey, almost complete to $K = 11.25$, have been described by \citet{2006MNRAS.368.1515E, 2006MNRAS.373...45E} and \citet{2007ApJ...655..790C}.  The collection of redshifts is continuing and it can be expected that deeper versions of the catalog will be made available in due course.

\begin{figure}
\figurenum{1}
\centering
\includegraphics[scale=0.4]{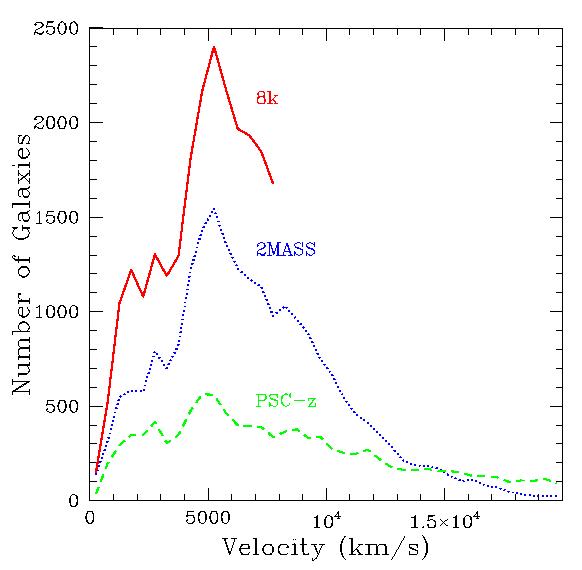}
\caption{Comparison of three all-sky redshift catalogs.  {\it V8k} (solid red curve) is compiled from a literature search.  The 2MASS survey (dotted blue curve) is an almost complete sample of galaxies with 2 micron fluxes $K<11.25$.  The PSC-z sample (dashed green curve) represents galaxies detected with IRAS brighter than 0.6~Jy at 60~microns.}
\label{8k_2mass_pscz}
\end{figure}

The {\it V8k} catalog was compiled from data from the literature, drawing heavily from J. Huchra's ZCAT\footnote{http://cfa-www.harvard.edu/$\sim$huchra/zcat} circa 2003.  The catalog explicitly excludes information from the large multi-object spectroscopy Sloan Digital Sky Survey (SDSS) and  Two Degree Field (2dF) survey.  The interest here is to have as uniform coverage as possible around the entire sky.  The SDSS and 2dF information is strongly directional.  The motivation for the accumulation of {\it V8k} was to have an overview of the distribution of nearby galaxies driven by some outreach activities.  The {\it V8k} database has become incorporated in the popular software packages Starry Night\footnote{http://www.starrynightstore.com} and Partiview\footnote{http://haydenplanetarium.org/universe/partiview} and are the basis of planetarium presentations\footnote{http://www.skyskan.com}.   In Figure~\ref{8k_2mass_pscz} there is a comparison of the numbers of galaxies with distance in {\it V8k}, {\it 2MASS $K<11.25$}, and PSC-z, the redshift survey based on the Infrared Astronomical Satellite (IRAS) Point Source Catalog  \citep{2000MNRAS.317...55S}.  The 2MASS catalog is more rigorously defined but the {\it V8k} catalog is denser. It contains low surface brightness and HI rich systems that do not register in 2MASS.

\begin{figure} 
\figurenum{2}
\centering
\includegraphics[scale=1.15]{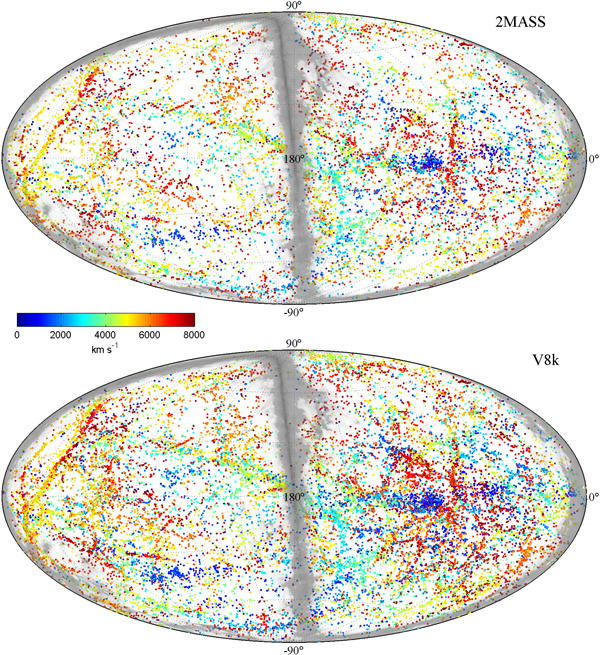}
\caption{Supergalactic coordinate projections of the distribution of galaxies in the {\it 2MASS $K<11.25$} and {\it V8k} catalogs.  Colors indicate redshift from $V \sim 0$ (blue) to $V \sim 8000$~\kms\ (red).  The {\it V8k} catalog is noticeably denser.  The region of obscuration in the Galactic plane is shown in grey.  There is still substantial incompletion at $\vert b \vert < 5\degr$.}
\label{v8k_aitoff}
\end{figure}

There are outstanding plots of the distribution of the 2MASS sample in \citet{2006MNRAS.373...45E}.  Figure~\ref{v8k_aitoff} displays the distribution of the {\it V8k} sample and compares it with  the {\it 2MASS $K<11.25$} sample over the same velocity range.  There are comparable numbers of galaxies in the {\it 2MASS $K<11.25$} and {\it V8k} catalogs but the former extends to greater depth and the latter gives more complete local coverage.

\begin{figure} 
\figurenum{3}
\centering
\includegraphics[scale=1.1]{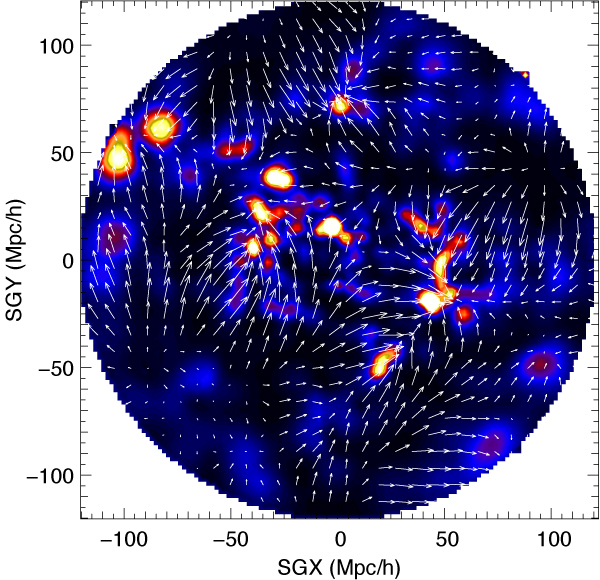}
\caption{Thin slice of the 2MASS Redshift Survey density field in the supergalactic equatorial plane.  We are at one edge of a supercluster complex extending $\sim 60 h^{-1}$~Mpc into the upper left quadrant where $h = {\rm H}_0 / 100$ and H$_0$ is the Hubble Constant.  The structure at the upper left corner is a bit of the massive Shapley Concentration.  The Perseus--Pisces filament is at the lower right and the Coma Cluster within the Great Wall is at upper center.  The arrows give a solution for the velocity field based on an orbit reconstruction.}
\label{lavaux_dens_vel}
\end{figure}

The redshift surveys are a necessary input for dynamical studies such as that by \citet{2006MNRAS.373...45E}.  One needs information on completion.  Assumptions are to be made regarding the relationship between visible objects and mass.  Distances to a fraction of the galaxies provide constraints on the cosmological models and the mass--light relationship \citep{1995ApJ...454...15S, 2005ApJ...635L.113M}.  Figure~\ref{lavaux_dens_vel} shows the result of a reconstruction of the {\it 2MASS $K<11.25$} sample  \citep{2008arXiv0810.3658L}.  This brief discussion serves just to situate the important role of the redshift catalogs in EDD.

\begin{figure} 
\figurenum{4}
\centering
\includegraphics[scale=0.55]{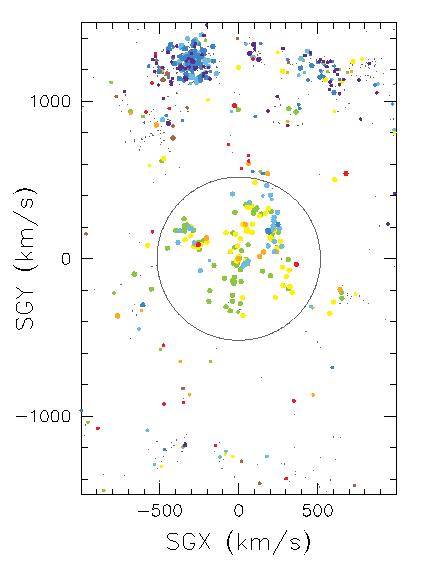}
\caption{Peculiar velocities in a 3 Mpc slice in supergalactic coordinates.  The Galaxy is at the origin of the coordinate system and the circle indicates a radius of 7~Mpc.  The Virgo Cluster is the clump of objects at the upper left.  After accounting for a cosmic expansion of 74~\kmsMpc, yellow-orange-red hews indicate peculiar velocities away from us and green-blue-purple hews indicate peculiar motions toward us.  Peculiar velocities within 7~Mpc are small (yellow, green: $<100$~\kms).  The local region has a bulk peculiar velocity toward Virgo of $\sim200$~\kms\  \citep{2008ApJ...676..184T}. }
\label{locsheet}
\end{figure}

The {\bf Catalog of Neighboring Galaxies} \citep{2004AJ....127.2031K}, found in the first block, is an all-sky compilation of galaxies with velocities with respect to the Local Group, $V_{LG}$, less than 550~\kms\ or distance estimates less than 10~Mpc.  Most of the entries are dwarfs.  Many of these objects have distances that are very well determined.  Those available to Karachentsev et al. at the time of their publication are recorded in this catalog.  A very important component of EDD is distance determinations based on the Tip of the Red Giant Branch (TRGB) method  \citep{1993ApJ...417..553L}.  A goal of this program is to make appropriate observations of as many galaxies in the {\it Catalog of Neighboring Galaxies} as possible.  Color--magnitude diagrams and TRGB distances will be discussed in a later section.  Currently, roughly half the galaxies in the {\it Catalog of Neighboring Galaxies} have TRGB distance measures.  With distances one can infer line-of-sight peculiar velocities and Figure~\ref{locsheet} is a demonstration of our current knowledge.  The circle in this plot gives the outer projection of a 7 Mpc sphere center at our position.  There is now a high level of completion of distance measurements within this region.

 The two afore mentioned redshift catalogs extend beyond 100~Mpc while the {\it Catalog of Neighboring Galaxies} is restricted to 10~Mpc.  The catalog called {\bf Tully 3000} is intermediate.  The table given here is an extension of the Nearby Galaxies (NBG) catalog \citep{1988ngc..book.....T}.  The limit of $V_{LG} = 3000$~\kms\ is the same as with NBG catalog but the number of entries is $\sim50\%$ greater.  The new catalog is still quite incomplete: {\it V8k} contains twice as many galaxies within 3000~\kms.  However, {\it Tully 3000} has two value-added features.  First, it provides specific information used to derive luminosity--linewidth distances (magnitudes in various passbands, inclinations, and linewidths) and gives distances as discussed in connection with the catalog {\it Tully08 Distances}.  Second, the table provides continuity with NBG catalog with  the assignment of galaxies to groups, down to groups of one \citep{1987ApJ...321..280T}.  The first 32 columns in {\it Tully 3000} give information specific to each individual galaxy.  The last 19 columns give information for the group containing the galaxy, such as total luminosity, mean velocity and, most important, a weighted mean distance.  These last 19 columns repeat the same information for every member of a specific group.
 
There are three entries in the first block relating to the {\bf Flat Galaxy Catalog} \citep{1999BSAO...47....5K}.  This catalog has special application to luminosity--linewidth studies because the selection of extremely thin galaxies assures the targets are spirals near class Sc and eliminates inclination ambiguity.  The selection rules are so restrictive that local coverage is sparse but the catalog provides a good sample for low density coverage of the range 3,000--10,000~\kms.  Two of the catalogs provide results.  See \citet{2000BSAO...50....5K} in one case.  In the other that makes use of 2MASS photometry, the table was provided privately by I. Karachentsev and discussed by  \citet{2002A&A...396..431K}.

A table in the first block is identified as {\bf Saunders PSC-z} because it is drawn from the IRAS Point Source Catalog 0.6 Jy redshift survey \citep{2000MNRAS.317...55S}.  It contains objects extracted from PSC-z that are targets for an observing program.  Like the flat galaxies, the PSC-z sources have interesting properties as targets for distance measurements.  The objects in {\it Saunders PSC-z} are temperature selected on the basis of $60-100~\mu{\rm m}$ color to isolate a normal spiral population,  with the flux arising from metal-enriched interstellar dust.  The specific sample is restricted to $i>45\degr$ and $V<6000$~\kms.  It is uniformly selected as a function of declination.  It gives good penetration of the zone of obscuration although source identifications become difficult to impossible at very low latitudes.   We have ongoing photometry and spectroscopic programs of this sample.

The table called {\bf V3k MK$<$-21} presents another sample for which we are acquiring distances.  It is limited in velocity to $V_{LS} < 3000$~\kms, in type by the exclusion of galaxies earlier than Sa, in inclination to $i>45\degr$, and in intrinsic luminosity by the limit $M_K < -21$.  The $K$ magnitude is from 2MASS and the translation from apparent to absolute magnitude uses distances based on a Numerical Action Model \citep{1995ApJ...454...15S} constrained by the measured distances reported in {\it Tully 3000} \citep{2008ApJ...676..184T}.  Within these limiting criteria the sample should be almost complete which minimizes concerns about Malmquist bias.  Selection by $K$ magnitude minimizes bias against edge-on systems and maximizes latitude coverage.

The final table in the first block is the {\bf Virgo Cluster Catalog} of \citet{1985AJ.....90.1681B}.  Presently the only value-added features to the published table are the PGC links, names, and J2000 coordinates.  Look in the future for the addition of multi-color magnitudes derived from SDSS images. 

\subsection{Summary Distances}

Presently there are only two catalogs in the second block but this number will grow.  This space is reserved for catalogs that accumulate distance estimates from a variety of sources and may assimilate as well as accumulate.  

{\bf Quality Distances} only accumulates.  It is a tabulation of distance moduli from the literature based on the following methodologies: the Cepheid Period--Luminosity Relation (PLR), the Tip of the Red Giant Branch (TRGB) method,   Surface Brightness Fluctuations (SBF), and in the miscellaneous column such other methods with limited application like those based on RR Lyrae pulsations, eclipsing binaries, and the maser observations of NGC~4258.    If there are multiple publications on a given galaxy by the same method, only one of these is carried into the {\it Quality Distances} catalog.  We do not give uncertainties but the authors typically claim $1 \sigma$ accuracy of 10\% or better for an individual measurement with these high quality techniques.  In a separate column there is a subjective average of distance moduli determined by different methods.

\begin{figure} 
\figurenum{5}
\centering
\includegraphics[scale=0.4]{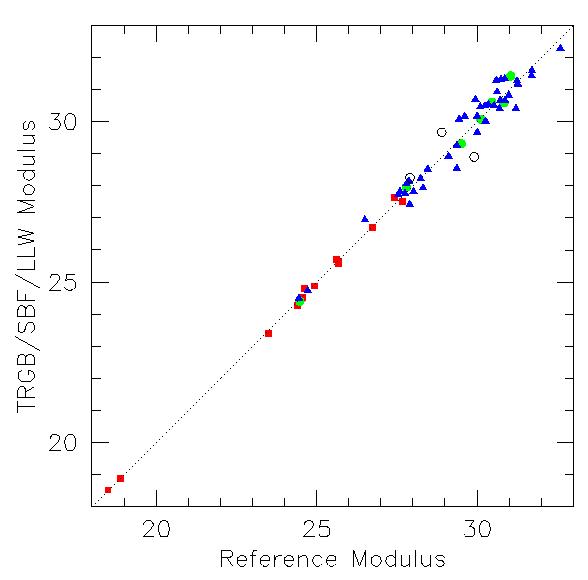}
\caption{Comparison of distance modulus measurements by different techniques.  In the cases of the red squares (TRGB) and the green filled circles (SBF) the reference moduli on the horizontal axis is given by Cepheid PLR measurements. The remainder are luminosity--linewidth (LLW) measurements, either
those from the Karachentsev et al. {\it Flat Galaxy Catalog} with 2MASS photometry (black open circles) or those from this program (blue triangles), compared with Cepheid PLR or TRGB on the reference axis.
See \citet{2008ApJ...676..184T} for details.}
\label{compare_dm}
\end{figure}

The {\bf Tully08 Distances} catalog gives an accumulation and assimilation of distance determinations for galaxies with $V_{LS} < 3000$~\kms\ as of the publication of  \citet{2008ApJ...676..184T}\footnote{LS = Local Sheet.  This alternative reference frame to LG = Local Group is defined by \citet{2008ApJ...676..184T}}.  The Hubble Space Telescope (HST) Distance Scale Key Project observations of the Cepheid PLR \citep{2001ApJ...553...47F} provides the scale zero point.  TRGB \citep{2007ApJ...661..815R}   and SBF \citep{2001ApJ...546..681T} measurements are confirmed to be on a consistent scale.  Finally our luminosity--linewidth distances are determined to be consistent.  A summary of comparisons is provided in Figure~\ref{compare_dm}.

The catalog {\it Tully08 Distances} combines tables 1 and 2 from \citet{2008ApJ...676..184T}.  The former table pertains to individual galaxies and  the latter to groups.  The catalog combines these elements in a similar fashion as seen in {\it Tully 3000} in the first block.  Group averaged information is appended to the entries for each of the group members.   The major difference between the catalogs {\it Tully 3000} and {\it Tully08 Distances} is that inclusion in the former is on the basis of redshift while the latter is restricted to the subset with well measured distances.

\subsection{Miscellaneous Distances}   
 
 Currently three distinct and important methodologies are represented in catalogs in the third block.  The first, {\bf CMDs/TRGB}, is the entry point into one of the most important components of EDD.  It is described in detail in a companion article \citep{2009unpublished_edd_cmd}.  A summary is given here.
 
\begin{figure} 
\figurenum{6}
\centering
\includegraphics[scale=0.6]{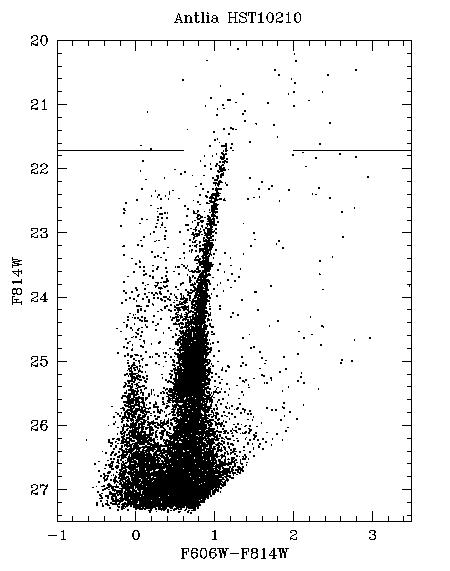}
\caption{Color--magnitude diagram for Antlia dwarf galaxy.  Stellar photometry data comes from observations with HST ACS.  The F814W filter approximates Cousins $I$ and the F606W filter approximates Johnson $V$.  The magnitude of the TRGB is indicated by the broken horizontal line.}
\label{antlia}
\end{figure}

The Tip of the Red Giant Branch (TRGB) method is emerging as arguably the best way to get distances to nearby galaxies, especially since the availability of Advanced Camera for Surveys (ACS) on HST.  The Red Giant Branch is well developed in stellar populations older than 2 Gyr (ie, in almost all galaxies) and stars at the tip are easily resolved with ACS in a single HST orbit if the distance is less than 10~Mpc   \citep{1993ApJ...417..553L, 1996ApJ...461..713S, 2002AJ....124..213M, 2006AJ....131.1361K, 2006AJ....132.2729M, 2007ApJ...661..815R}.

An example of a Color--Magnitude Diagram (CMD) obtained of the galaxy Antlia is shown in Figure~\ref{antlia}.  This CMD preserves a record of Antlia's star formation history that is rich in detail.  The Red Giant Branch is particularly prominent.  Stars advance up the Red Giant Branch during the phase when Hydrogen burning is occurring in a shell around a Helium core supported by electron degeneracy.  The degeneracy is broken with the onset of Helium burning at a well defined core mass, accounting for the termination of the Red Giant Branch at a distinctive luminosity.  Age and metallicity conditions affect the envelope of the star.  In older and more metal rich stars the photospheres are cooler.  At blue bands the brightest RGB stars are young or metal--poor, while at infrared bands the brightest RGB stars are older or metal--rich.  Empirically it is fortuitously found that at $I$ band, or the F814W filter in HST ACS, there is very little dependency of the TRGB on age or metallicity.  The TRGB is an outstanding standard candle at this wavelength.

HST ACS and the predecessor Wide Field--Planetary Camera 2 (WFPC2) have been used to observe the resolved stellar populations of nearby galaxies over many HST cycles and by many Principal Investigators.  Currently there are data for $\sim 250$ galaxies in the archive.  A large fraction of the total were observed specifically for the purpose of measuring distances.  References for the early work are found in the {\it Catalog of Neighboring Galaxies} in the first block.  More recent results are reported by \citet{2006AJ....131.1361K} and \citet{2006AJ....132..729T}.

The HST observations, whether with WFPC2 or ACS, are carried out in two filters that approximate $V$ (either F555W or F606W) and $I$ (F814W).  The stellar photometry is done with the package HSTPHOT with WFPC2 images \citep{2000PASP..112.1383D} and the extension of that program called DOLPHOT\footnote{http://purcell.as.arizona.edu/dolphot} with ACS images.  The program allows for the creation and resampling of artificial stars to monitor issues of completion with crowding and faintness.  Our maximum likelihood procedures for measuring the TRGB are described by \citet{2006AJ....132.2729M} and calibration topics are discussed by \citet{2007ApJ...661..815R}.

The database in {\it CMDs/TRGB} gathers all the relevant observations made with HST WFPC2 or ACS by all programs but as analyzed by our group with the HSTPHOT and DOLPHOT photometry programs, our tip finding algorithms, and our calibrations.  The results are presented in tabular form in the catalog {\it CMDs/TRGB}.  Color--magnitude diagrams, with TRGB fits superimposed, HST footprints, galaxy images, and files of the stellar photometry can be accessed for individual galaxies within {\it CMDs/TRGB} \citep{2009unpublished_edd_cmd}.

The other two catalogs in the Miscellaneous block are based on tables taken directly from the literature.  {\bf Tonry SNIa} is Table 15 from \citet{2003ApJ...594....1T}.  {\bf Tonry SBF} is Table 1 from \citet{2001ApJ...546..681T}.  The only value--added feature in these two cases is the PGC name linkage.  The SBF distances have been confirmed to be consistent in scale with the HST Key Project standard and have been incorporated into the assimilation catalog {\it Tully08 Distances} discussed in connection with the second block.  The SNIa distances have not yet been confirmed to be on the same scale.  Giving attention to that matter is one of the most important things we still have to do.

\subsection{Photometry}

This block contains a large number of catalogs although some are only of historical interest.  The catalog {\it Hawaii Photometry} presents new material in both tabular and graphic forms.  We will return to discuss this important contribution after a brief discussion of the rest of the block.

\begin{deluxetable*}{lrccll}
\tablenum{1}
\tablecaption{Comparisons Between Alternate Sources of I Band Photometry}
\label{tbl:Iphot}
\tablewidth{0in}
\tablehead{\colhead{Source} & \colhead{Offset} & \colhead{St. Dev.} & \colhead{No.} & \colhead{Outer} & \colhead{Reference}}
\startdata
Mathewson   &   $0.00$  &            &    -     & Tot to sky & \citet{1996ApJS..107...97M} \\
Haynes          & $-0.01$ & 0.006 &  287 & 8$h$           & \citet{1999AJ....117.1668H} \\
Han/Mould    & $+0.10$ & 0.010 & 199 & $\infty$        &   \citet{1993ApJ...409...14M}\\
Pierce             &   $0.00$  & 0.011 & 150 & 24/as$^2$ &   \citet{2000ApJ...533..744T}\\
Roth                & $-0.06$  & 0.011 &   56 & 23.5/as$^2$ &  \citet{1994AJ....108..862R} \\
Dale                & $-0.01$  & 0.014 &   50 & 8$h$            &   \citet{1999AJ....118.1468D}\\
Dell'Antonio   & $+0.12$ & 0.022 &   41 & $\infty$         &  \citet{1996AJ....112.1759D}\\
Verheijen       &   $0.00$  & 0.019 &   40 & $\infty$         &  \citet{1996AJ....112.2471T} \\
Schommer      & $-0.04$  & 0.018 &   31 & 24/as$^2$   & \citet{1993AJ....105...97S}  \\
Bernstein       &  $+0.06$  & 0.015 &  24 & $\infty$         &  \citet{1994AJ....107.1962B} \\
Bureau           &  $+0.07$ & 0.038 &   11 & $\infty$         &  \citet{1996ApJ...463...60B} \\
\enddata
\end{deluxetable*}

The catalog {\bf Homogenized Photometry} was built from an assimilation of the original photometry from the other catalogs.  The most important contributions come from {\bf Mathewson}  \citep{1996ApJS..107...97M} and  {\bf Haynes SFI/SCI}  \citep{1999AJ....117.1668H}.  A comparison involving 287 galaxies observed by both teams reveals a tiny offset with $2\sigma$ significance between the two: $I_{math} = I_{hay} - 0.01$.  This small difference might be expected since the {\it Mathewson} magnitudes are `total to sky' while the {\it Haynes} magnitudes are to a radius of 8 exponential scale lengths, $h$.
We accept the {\it Mathewson} zero point and subtract 0.01 mag from the Haynes et al. values.  Then we accept the {\it Mathewson} and {\it Haynes} results as equal and use their magnitudes as the $I$ band standards against which all the other sources are gauged.  Comparisons between sources are shown in Table~\ref{tbl:Iphot}.   The zero point offsets are in the sense `X + offset = standard' where X is the source under consideration and `standard' is  the average of the {\it Mathewson} and {\it Haynes} magnitudes.  

The $I$ photometry from the sources identified in Table~\ref{tbl:Iphot} were averaged to produce the magnitude entries in the catalogs {\it Tully 3000} and {\it Tully08 Distances}.  In cases where there are multiple sources, strongly deviant values were filtered out.  Bad data were easily identified if there are 3 or more independent measurements and can be reliably identified with two independent measurements and/or color information.

There is more limited information at other photometric bands.  Our data at $B$ band all comes from the program carried out by Pierce and Tully during the 1980's and '90s.  The $R$ band material comes from the Pierce and Tully program and also from programs by \citet{1997ApJS..109..333W} and \citet{1997AJ....114.2402C}.  In those latter cases the observations were in Gunn $r$ band and were transformed to Cousins $R$ with:
$$R = r_{gunn} - 0.39 + 0.28 (R-I) - 0.05 A_B^b$$
and if no color information we assume $R-I = 0.56$.  Using the Pierce and Tully data as the $R$ standard, after these transformations we find:
$$R_{will} -R_{stand} = 0.16 \pm 0.031~~25~{\rm cases}$$ 
$$R_{cour} -R_{stand} = 0.14 \pm 0.062~~11~{\rm cases}$$ 

The photometry programs also give estimates of inclinations through measurements of galaxy ellipticities.  The inclination from face-on, $i$, is related to the ratio of the dimensions of the major axis, $a$ and minor axis, $b$ by the formula:

\begin{eqnarray}
{\rm cos}~i = \sqrt{{(b/a)^2 - q_0^2}\over{1 - q_0^2}}
\label{inc}
\end{eqnarray}

The parameter $q_0$ represents the intrinsic axial ratio of a galaxy seen edge-on.  It  can be argued that the value adopted for $q_0$ should depend on galaxy morphology \citep{1997AJ....113...22G}.  However this involves the requirement of additional information and we have found the most important concern for the avoidance of systematics is consistency.  We adopt $q_0 = 0.20$ independent of type.  

Values of $b/a$ have been accumulated and inter-compared from the sources identified in Table~\ref{tbl:Iphot}.   Results are very consistent between sources.  The only significant deviation is with  \citet{1996ApJS..107...97M}.  We find:
$$i_{math} - i_{other} = -2.0 \pm 4.4 (\pm 0.3~{\rm st.dev.)}$$
from 274 cases.  Consequently we adjust the {\it Mathewson} inclinations by $+2\degr$.  Agreement between the other sources is within $2\degr$ in the mean, usually within $1\degr$.  Comparisons were
made between literature sources and between filter passbands and instances of large deviations were rejected.    The results of the comparisons made there way into the catalogs {\it Tully 3000} and {\it Tully08 Distances}. 

The catalog {\bf McDonald Virgo} is based on recent observations on the University of Hawaii 2.2m Telescope, the Canada-France Hawaii Telescope, and the United Kingdom Infrared Telescope at Mauna Kea Observatory.  A complete magnitude-limited sample of galaxies in the Virgo Cluster have been observed in $H$ band \citep{2009arXiv0901.3554M}.

{\bf Cornell Photometry} \citep{2007ApJS..172..599S} is another synthesis catalog.  It appeared after the compilation of {\it Homogenized Photometry} so the two big assimilations represent separate compilations of literature data.  The original information is often the same.

We return to the {\bf Hawaii Photometry} catalog.  As in the case of {\it CMDs/TRGB}, this catalog provides tabular and graphic information resulting from our own observations and the full story is told in a separate paper in this series \citep{2009unpublished_edd_photometry}.

The observations contributing to the {\it Hawaii Photometry} catalog were made between 2000 and 2008 with the University of Hawaii 2.2m telescope equiped with a Tek2048 CCD at the f/10 Cassegrain focus.  The field of view is $7.5^{\prime}$ and pixel scale is $0.22^{\prime\prime}$.  Most observations were 300s in Cousins $I$ band.  A small fraction of targets were also observed in Cousins $R$ and Johnson $B$ bands for 300s and 600s respectively.

\begin{figure} 
\figurenum{7}
\centering
\includegraphics[scale=0.4]{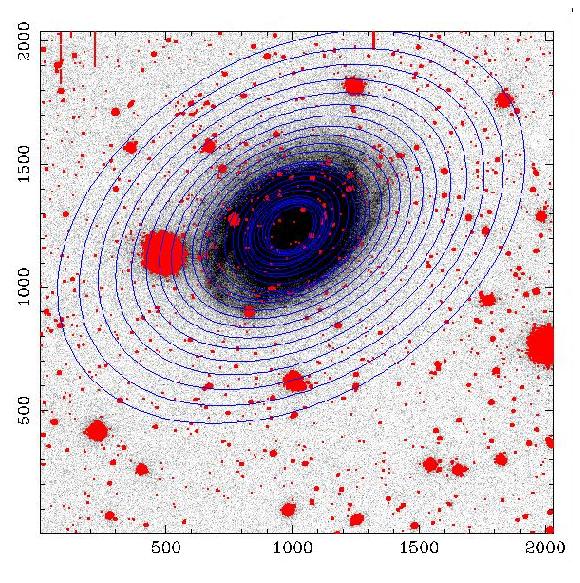}
\includegraphics[scale=0.4]{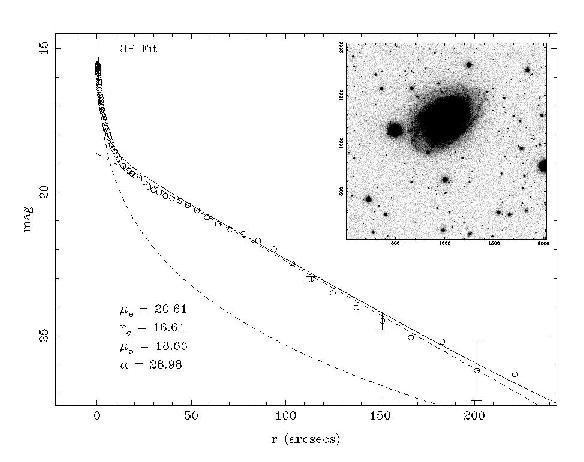}
\includegraphics[scale=0.4]{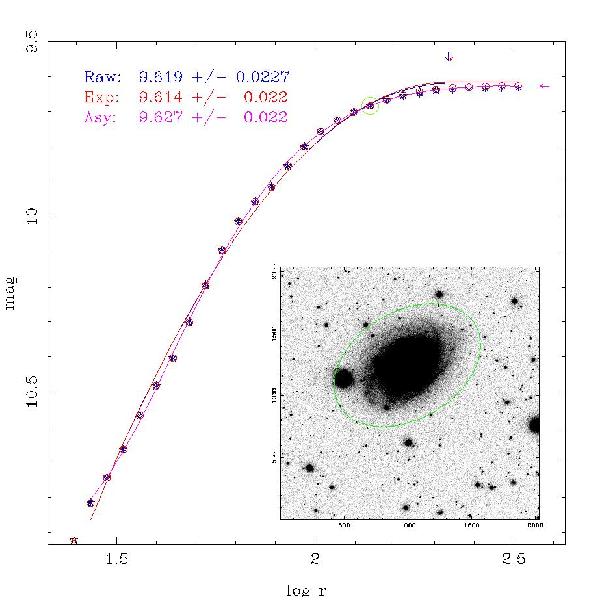}
\caption{Photometry of NGC 3223.  Graphic output from the Archangel photometry program.  Top panel: galaxy image with superposed bad pixel map and ellipse fit.  Middle panel: surface photometry based on ellipse fits; curves based on 2-component (exponential and $r^{1/4}$) surface brightness models.  Bottom panel: magnitude growth curve. }
\label{photometry}
\end{figure}

The data analysis was carried out with the photometry software package Archangel\footnote{http://abyss.u.oregon.edu/$\sim$js/archangel}.  The program carries out star masking, ellipse fitting, then the compression to one--dimensional information such as the run of ellipticity, surface brightness, and total magnitude with radius.  Extrapolations that are usually very small are made to derive total magnitudes.  An example of graphical data products are shown in Figure~\ref{photometry}.  The catalog {\it Hawaii Photometry} is a tabulation of the parameters characterizing each target and the access point to graphic material such as seen in Fig.~\ref{photometry}.

\subsection{HI Linewidths}

This block presents a combination of new and literature data pertaining to neutral Hydrogen line profiles.  The new material is contained in catalog {\it All Digital HI}.  Before introducing that, we first discuss the catalogs assembled from other sources.

{\bf Pre-Digital HI} gives a table of information from years before digital archives that was compiled by the lead author and Cyrus Hall.  The sample is restricted to velocities less than 3,000~\kms.  The parameter of greatest interest in this table is the HI linewidth estimator $W_{20}$, the width at 20\% of peak intensity.  This parameter was measured with a (very small) ruler from paper copies of HI profiles. The procedure was arduous and not as quantifiable as modern algorithmic methods but for profiles with peak signal above 7 times noise the measurements are made consistently.  This catalog provides a basis for comparisons with computer analyzed data and for very large, nearby galaxies may represent the only available information.

The catalog {\bf Springob/Cornell HI} \citep{2005ApJS..160..149S} is a very big and important collection of HI data.  It is available alternatively through a Cornell website\footnote{http://arecibo.tc.cornell.edu/hiarchive} or through the NASA/IPAC Extragalactic Database (NED)\footnote{http://nedwww.ipac.caltech.edu/forms/SearchSpectra.html}.  The compendium includes data obtained by the Cornell group and their collaborators with Arecibo 305m, Green Bank 92m and 43m, Nancay 200x40m and Effelsberg 100m telescopes.  Only one profile is shown for a given galaxy -- presumably the best one available.  The digital profile was processed through a pipeline developed by the Cornell group.  Among the products are five alternative linewidth measures,  In addition to the tabulated data, the catalog {\it Springob/Cornell HI} provides links to profile displays at the Cornell web site.

The catalogs {\bf HI Nancay}\footnote{http://klun.obs-nancay.fr}  \citep{2006ASPC..351..429T}, and {\bf HIPASS 1000} \citep{2004AJ....128...16K}, and {\bf WHISP}\footnote{http://www.astro.rug.nl/$\sim$whisp}   \citep{1996A&AS..116...15K} provide tabulated and graphic information drawn from literature sources.  In each case, the unique feature with EDD is the easy linkage between sources.  The catalog {\bf HI Fisher} gives web links to pre-publication material at a website maintained by J.R. Fisher\footnote{http://www.cv.nrao.edu/$\sim$rfisher}.  This material is incorporated in the catalog to be discussed next.

The {\bf All Digital HI} catalog is a collection of data from new observations with the 100m Green Bank Telescope (GBT) and the 305m Arecibo Telescope and from the literature -- all analyzed in the same way.  The material from Arecibo Telescope was acquired after the installation of the Gregorian feed and ground screen but before the commissioning of the multifeed.  The results of initial reductions of this data are accessed through {\it HI Fisher}.  The GBT program is more extensive, begun in 2001 with early results reported in {\it HI Fisher} and ongoing.   Beginning in the fall of 2008, the observations have
continued as a Green Bank Telescope Large Proposal\footnote{http://www.vla.nrao.edu/astro/prop/largeprop/}.

These new data are combined with all archival material that could be found in {\it All Digital HI}.  The analysis program developed to interpret the newly acquired HI spectra was also applied to the archival spectra.  The brief description here is expanded upon by \citet{2009unpublished_edd_HI}.

The most important parameter for distance work provided in {\it All Digital HI} is the measure of linewidth, $W_{m50}$, the width of the HI profile at 50\% of the mean flux within the velocity range encompassing 90\% of the total HI flux.  The quality index that indicates whether the linewidth is acceptable for the purpose of a distance estimate is built into the error figure, $e_W$.  Acceptable profiles have $e_W \leq 20$~\kms.  Unacceptable profiles, for whatever reason, are assigned $e_W > 20$~\kms.

\begin{figure}[htb!]
\figurenum{8}
\centering
\includegraphics[scale=0.45]{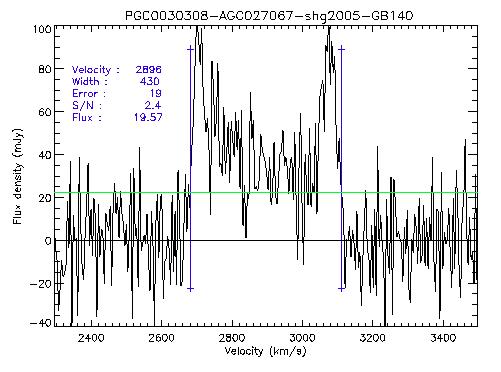}
\caption{The HI profile of NGC 3223 observed with the Green Bank 140 Foot Telescope.  The profile is drawn from the Cornell archive.}
\label{HIprofile}
\end{figure}

The tabular part of {\it All Digital HI} gives the results of the analysis of both the new observations and the old material from the archives.  Selection on a name gives access to graphical representations of the profile fit.  An example is given in Figure~\ref{HIprofile}

\subsection{Optical Linewidths}

This block reports results from two studies that attempt to reconcile optical rotation curve and HI linewidth measures of the dynamical state of galaxies.  The {\bf Catinella/Cornell} catalog \citep{2005AJ....130.1037C} gives the detailed results of comparisons between optical and radio kinematic information acquired by the Cornell group.  Then there are four catalogs identified with {\bf Courteau}, distinguished by the four sources of data used in the study of \citet{2007ApJ...671..203C}.  For the moment these tables from the literature are given without interpretation or evaluation.

\subsection{Fundamental Plane}

This currently the last block in EDD contains results from three major surveys based on the Fundamental Plane (FP) relationship between luminosity, surface brightness, and velocity in elliptical and lenticular galaxies.  The three surveys go by the names EFAR, ENEAR, and SMAC (Streaming Motions of Abell Clusters).  There is very little overlap between the first two but there is a substantial overlap with each of these and SMAC.  

In the case of SMAC, three catalogs are provided in EDD.  {\bf FP: SMAC3} is drawn from  \citet{2001MNRAS.327..265H} and gives the observational material needed for the derivation of FP distances for 56 rich clusters of galaxies.  {\bf Hudson SMAC FP} is based on a file provided by M.J. Hudson in advance of publication and provides distances to the 56 clusters.  The catalog {\bf Blakeslee SMAC FP}   \citep{2001MNRAS.327.1004B, 2002MNRAS.330..443B}  is a synthesis of results on nearby galaxies in the SMAC program that overlap with the SBF project previously discussed \citep{2001ApJ...546..681T}.  The SMAC scale is tied to the HST Key Project scale though a comparison with the large SBF sample.

\begin{figure}[htb!]
\figurenum{9}
\centering
\includegraphics[scale=0.4]{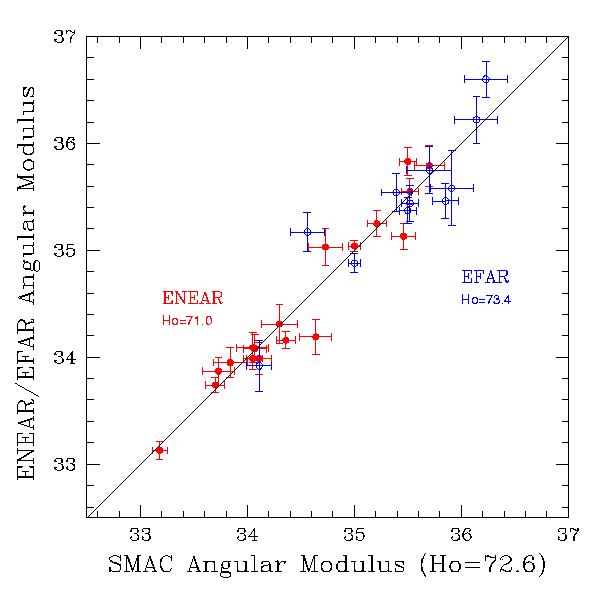}
\caption{Comparisons between three Fundamental Plane programs.  Angular distance moduli obtained with the SMAC program are compared with results by ENEAR (filled red symbols) and EFAR (open blue symbols).  The SMAC scale is set by comparison with SBF distances.  The ENEAR and EFAR scales are set to give optimal agreement with SMAC.  The three cases are consistent with a scale set by H$_0 = 72 \pm 1$~\kmsMpc.}
\label{smac_enear_efar}
\end{figure}

The ENEAR galaxies are mostly within 100 Mpc while the EFAR galaxies are mostly beyond this distance.  ENEAR is based on the variation of  the FP relation known as $D_n - \sigma$.  The sample contained in the catalog {\bf FP: ENEARc} is drawn from 28 clusters (hence the `c' ) \citep{2002AJ....123.2990B, 2002AJ....123.2159B}.  The catalog {\bf FP: EFAR} gives results for 81 clusters \citep{2001MNRAS.321..277C}.  The overlap in samples between SMAC and the other two surveys permits the comparison in results seen in Figure~\ref{smac_enear_efar}.  
 
\section{Summary}

EDD has emerged out of the personal need of the authors to find order in the chaos of information related to galaxy distances.  Within the context of a given methodology, it  can be necessary to pull together diverse observational components from a multitude of sources.  Then there is the need to make comparisons between methodologies.  Statistically significant comparisons have to be made in two matrix dimensions: on specified targets over many methodologies or sources and on specific methodologies or parameters over many targets.

In EDD, comparisons are enabled by the linkage of catalogs to the discrete naming convention of {\it LEDA}.  As examples, a user could look for all overlapping sources of $I$ band photometry or all overlapping distance measures whether by object or technique.  One can hone the comparison through use of the maximum/minimum and sort features.  Many of these kinds of comparisons have already been made (and are continuing) and synthesis tables appear in EDD and are updated as information improves.

Finally, and very importantly, EDD provides a way for the authors to make available observational material associated with three ongoing programs: one involving galaxy photometry resulting from CCD imaging  \citep{2009unpublished_edd_photometry}, another involving the measurement of HI profile widths from single dish radio telescope observations \citep{2009unpublished_edd_HI}, and the third involving the analysis of color--magnitude diagrams obtained from HST observations that reveal the resolved stellar populations in nearby galaxies \citep{2009unpublished_edd_cmd}.  With the latter two programs the material made available in EDD is both from our own observations and from archival data passed through our software pipelines.

\vskip 1cm
\noindent
The genesis of EDD was provided by the need for a database to manage the Space Interferometry Mission (SIM) Dynamics of Galaxies Key Project.  It was subsequently expanded to manage the Hubble Space Telescope (HST) programs AR-9950, GO-9771, 10210, 10235, and 10905.  D.I.M. acknowledges support from Russian Foundation for Basic Research grant 08-02-00627.  Tremendous use has been made of  NED, the NASA/IPAC Extragalactic Database operated by the Jet Propulsion Laboratory, California Institute for Technology, the HyperLeda database at the University of Lyon 1, and the Center for Astrophysics Redshift Survey website maintained by John Huchra.  By it's nature, the database derives most of its content from the literature and other archives.  We are grateful to all.

\bibliography{paper}
\bibliographystyle{apj}

\end{document}